\documentclass[preprint,12pt,authoryear]{elsarticle}
\usepackage[margin=1in]{geometry}
\usepackage{amsmath,amssymb,amsthm,amsfonts,graphics,upgreek,babel,geometry,array}
\allowdisplaybreaks
\usepackage[dvipsnames]{xcolor}
\usepackage{ragged2e}
\usepackage{microtype}
\usepackage{float}
\usepackage{graphicx}
\usepackage{longtable} 
\usepackage{ulem}
\usepackage{booktabs}
\usepackage{multirow}
\usepackage{tabularx}
\usepackage[official]{eurosym}

\usepackage[colorlinks=true, allcolors=blue]{hyperref}
\usepackage{lipsum} 
\usepackage{natbib}
\usepackage{float}
\usepackage{caption} 
\usepackage{booktabs} 

\date{}

\journal{ }

\begin{document}

\begin{frontmatter}

\author[DIME]{Federico Gallo}
\author[DIME]{Alireza Shahedi}
\author[DIME]{Angela Di Febbraro}
\author[DELFT]{Mahnam Saeednia}
\author[DIME]{Nicola Sacco\corref{cor1}}
\ead{nicola.sacco@unige.it}

\affiliation[DIME]{organization={Dept. of Mechanical, Energetic, Management, and Transport Engineering, University of Genoa},
           addressline={Via Montallegro 1}, 
           city={Genoa},
           postcode={16145}, 
           country={Italy}}
            
\affiliation[DELFT]{organization={Dept. of Transport and Planning, Delft University of Technology},
           addressline={Postbus 5}, 
           city={Delft},
           postcode={2600 AA}, 
           country={The Netherlands}}

\cortext[cor1]{Corresponding author}

\title{Robustness Evaluation of a Physical Internet-based Intermodal Logistic Network} 

\begin{abstract}
\justifying
The Physical Internet (PI) paradigm, which has gained attention in research and academia in recent years, leverages advanced logistics and interconnected networks to revolutionize the way goods are transported and delivered, thereby enhancing efficiency, reducing costs and delays, and minimizing environmental impact. Within this system, \textit{PI-hubs} function similarly to cross-docks enabling the splitting of PI-containers into smaller \textit{modules} to be delivered through a network of interconnected hubs, allowing dynamic routing optimization and efficient consolidation of PI-containers. Nevertheless, the impact of the system parameters and of the relevant uncertainties on the performance of this innovative logistics framework is still unclear.  
For this reason, this work proposes a robustness analysis to understand how the PI logistic framework is affected by how PI-containers are handled, consolidated, and processed at the PI-hubs.
To this end, the considered PI logistic system is represented via a mathematical programming model that determines the best allocation of PI-containers in an intermodal setting with different transportation modes. In doing so, four Key Performance Indicators (KPIs) are separately considered to investigate different aspects of the PI system's performance and the relevant robustness is assessed with respect to the PI-hubs' processing times and the number of modules per PI-container. In particular, a Global Sensitivity Analysis (GSA) is considered to evaluate, by means of a case study, the individual relevance of each input parameter on the resulting performance.
\end{abstract}


\end{frontmatter}

\section{Introduction}
Rail-road intermodal transportation has emerged as one of the crucial solutions to enhance the efficiency and sustainability of modern logistic networks. It has attracted the attention of researchers during the past decades to address the relevant economic, social, and environmental issues. In this connection, on one hand the dependency between road freight and the socioeconomic systems was proved (\cite{CARRARA2017359}). On the other hand, it has clearly emerged the need to reducing traffic congestion and shifting freights from road transportation to environmentally friendly ones (e.g. rail transportation) to reduce the negative environmental impacts (\cite{turnbull2015role}).
These motivations are driving carriers and terminals to organize road transportation at the best and integrate with the other transportation modes to achieve an optimized and more sustainable freight delivery (\cite{sun2018multiagent}). 

Road freight transportation contributes to greenhouse gas emissions and is the larger emitter of greenhouse gas generator among the different transportation options (\cite{World2017}). Therefore, one of the main strategies for improving sustainability is freight transportation optimization, with a particular focus on shifting the transportation process from road to rail. 
The Physical Internet (PI), proposed by Montreuil (2011), is an innovative logistics concept aimed at enhancing efficiency and sustainability in global supply chains. By adopting open, modular, and standardized systems, PI promotes seamless collaboration, better resource utilization, and reduced environmental impacts, driven by intelligent integration and optimized decision-making across interconnected networks (\cite{montreuil2011toward, Piroadmap}).
The so-called Physical Internet (PI) is a solid example of digital transformation of the supply chain operations, as it integrates advanced technologies such as real-time tracking, automation, and data analytics to optimize logistics, streamline transportation, and enhance overall operational efficiency. By addressing inefficiencies through digital tools, the PI reduces the environmental impact of logistics operations, including pollution caused by heavy traffic. The main innovation in the logistics sector provided by PI could be from the freight transportation perspective, as it can optimize shipping operations, particularly in terms of capacity exploitation and reduction of the number of circulating trucks, which in turn reduces CO$_2$ emissions and traffic congestion (\cite{mckinnon2009innovation}). This is because freight operations play a crucial role in addressing sustainability goals and environmental challenges in logistics networks.

The growth of integrated intermodal transportation in the context of logistics and supply chain, including urban environment, is evident in the development of urban distribution systems and passenger transportation. The rail-road intermodal terminals are trying to optimize the traffic flow nowadays as the purpose of rail transit systems for regional mobility is to facilitate accessibility. 

Intermodal transportation can also promote shipping quality across modes on a global operations scale. Predictions indicate that the throughput of intermodal transportation is increasing (\cite{CARRARA2017359}), which creates favorable conditions for the adoption and growth of modular concepts in shipping and delivery logistics.

\begin{figure*}[h]
\centerline{\includegraphics[scale=0.9]{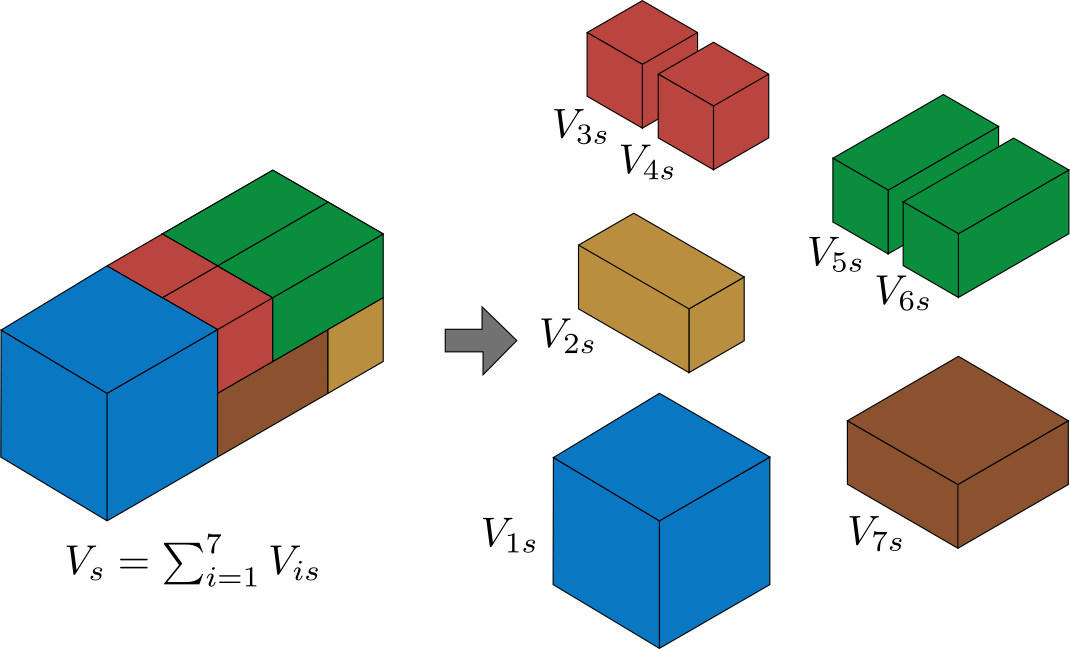}}
\caption{Example of a PI-container split into different modules with different volumes (\cite{shahedi2023lead}).}
\label{fig:modularShipment}
\end{figure*}

Furthermore, the PI framework is designed to integrate advanced automation technologies and intelligent systems, enabling dynamic decision-making and fostering real-time adaptability across interconnected logistics networks. One of the fundamental concepts of the PI is the modularization of a single PI-container, as depicted in Fig.~\ref{fig:modularShipment}. These modular PI-containers can be transported using various transportation modes and along different paths. This feature of the PI-containers in PI scheme allows for the exploitation of existing capacities in pre-planned service (e.g. trains or trucks) that do not have enough capacity to transfer the entire PI-container. The main advantage of this approach is the optimization of transportation mode capacities and the reduction of  the half-filled commuting vehicles. Nevertheless, the claim of a delivered PI-container can be made if and only if all the modules of that PI-container have been delivered to the destination.

As PI is a complex, distributed logistic network whose performances depends on effectiveness of the routing operation performed in intermodal nodes on a large set of modules, it is fundamental to identify the factors that may significantly affect its reliability and efficiency. To this end, based on the optimized PI scheme introduced in (\cite{shahedi2023lead}), this study proposes a robustness analysis of the system's capability to maintain performance under such uncertainties.
In particular, the goal of the proposed analysis consists of assessing how much the system's performances are affected by its PI-related parameters, such as the number of modules making up PI-containers, and the travel times of the processing times in the PI-hubs.
The results can help better understand the behavior of the PI logistic scheme and identify which parameters are most influential on the system performance and thus deserving deeper analyses or direct  inclusion in the optimization problems regulating the PI dynamics. 
 
\section{Literature Review}

This section reviews the main current knowledge on the Physical Internet concept (Section~\ref{lit:pi}) and the robustness evaluation methodologies (Section~\ref{lit:rob}). Finally, it summarizes the paper's main contributions (Section~\ref{lit:cont}).

\subsection{Physical Internet} \label{lit:pi}

Physical Internet is a novel concept for improving the efficiency of existing logistics systems from an automation perspective (\cite{montreuil2011toward}). In this context, PI is commonly defined as ``a global logistics system that connects networks using standardized rules, modular containers, and smart interfaces for better efficiency and sustainability'' (\cite{ballot2014physical}) which highlights logistics improvement based on ideas from the digital world.

PI gets its name from the conceptual similarities of this new logistical paradigm with the Digital Internet. In fact, PI performs in the same way as the Digital Internet, by following the same basic ideas of connectivity and flow. The DI is made up of a large network of servers and computers that send and receive data using standard rules and efficient pathways. In the same way, the PI uses PI-containers like ``data packets'' and PI-hubs like ``routers'' to move goods through connected networks (\cite{qiu2015physical, landschutzer2015containers, sternberg2021toward}).

In this framework, many studies have covered various aspects of PI, including the design of the required infrastructures (\cite{onal2018product}), technological innovation (\cite{treiblmaier2019combining}), and business models (\cite{montreuil2012physical}), preparing the ground for PI’s practical implementation. In this connection, the relevance of the concept is emphasized by the Alliance for Logistics Innovation through Collaboration in Europe (ALICE), which has defined and developed road-maps that highlight the vision for the global implementation of PI from 2020 to 2050 (\cite{Alice2017}). 

Regarding PI-containers transportation and delivery, many studies have tried to address the problems related to their optimal routing. In this framework, to strengthen the PI features of the global logistics system, \cite{chen2017using} presented an innovative solution for the container routing problem by considering taxis to collect e-commerce reverse flows in the city. \cite{chargui2019multi} proposed an optimization approach for a routing problem to transfer PI-containers between the cross-docks. Likewise, \cite{sallez2016activeness} considered a framework to address the routing of PI-containers in the context of cross-docks. Regarding the freight routing problem, \cite{puskas2019concepting} considered virtual transfer points to assess the possibility of reconfiguring platoons in the PI system. 

Scheduling and resource allocation in PI-hubs are issues that impact system performance, as demonstrated by extensive studies in the literature. Various works have addressed these challenges, such as scheduling problems, container allocation, and inventory management, using tools like queuing models, optimization methods, and inventory control models (\cite{karakostas2019modelling, zhang2020dynamic, kong2016scheduling, walha2016rail, chargui2020proposal, pan2015perspectives, yang2017innovative, yang2017mitigating}). However, since this study does not focus on these aspects, we will assume random performance values for PI-hubs and PI-containers.

\subsection{Global sensitivity analysis for model robustness evaluation} \label{lit:rob}

Robustness is a widely studied concept in statistics, with different definitions tailored to specific applications. In this context, GSA analysis was first introduced in (\cite{wagner1995global}), where an approach based on generating input-output Monte Carlo samples was proposed. These samples were analyzed using statistical techniques to estimate variance-based sensitivity indices, providing a systematic approach to evaluate the relative importance of input parameters. Since Wagner's foundational work, GSA methods have significantly diversified. For example, \cite{kleijnen1999statistical} explored nonparametric regression methods, whereas \cite{baucells2013invariant} developed moment-independent techniques. Other advancements include value-of-information techniques (\cite{felli1998sensitivity, strong2013efficient}) and the application of Shapley values to sensitivity analysis (\cite{owen2014sobol}).

Furthermore, GSA has been widely applied across engineering and applied sciences, including the transportation and logistics sector. For example, \cite{sacco_gsa} evaluated the performance robustness of different traffic light optimization approaches with respect to parameter uncertainty and variability, whereas \cite{gallo_gsa} conducted a GSA to identify the most relevant origin-destination pairs for planning demand-responsive railway services. Adidtionally, \cite{kioutsioukis2004uncertainty} presented a Monte Carlo-based methodology for assessing uncertainty in transport emission models. 
Regarding the aim of this work, GSA is considered to address the uncertainty that arising from the variability of the system inputs or parameters \cite{saltelli2008global}. In particular, when the system under analysis operates according to the solutions of optimization problems, the unpredicted variability in inputs and parameters can significantly degrade the expected performance if not adequately considered.
To address this issue, GSA assesses how inputs and parameters affect the output. By identifying a ``sensitivity pattern'', GSA ranks parameters based on their contribution to overall uncertainty, highlighting those that must be carefully calibrated or, where possible, explicitly incorporated into the optimization process, potentially through stochastic programming \cite{saltelli2004sensitivity}.

\subsection{Paper contributions} \label{lit:cont}

From the literature mentioned in Section~\ref{lit:pi}, it can be concluded that the concept of intermodal freight transportation in the PI sector has been extensively investigated. Also, the applicability of the PI in freight transportation, containers, terminals, and hubs has been investigated. 
However, it remains unclear how the PI key characteristics (i.e., the processing of the PI-containers at the PI-hubs and the splitting of the PI-containers into modules) may affect the performance of the freight transportation system, such as the delivery times or the transportation costs. This can be particularly important when uncertainties are present in the system. For example, the exact processing time for a PI-container at a PI-hub may not be known at the time the routing decision for the PI-container is made. Ignoring these aspects may result in degraded system performance of the system and in a decrease in the effectiveness of the optimized PI-container routing.

To cope with this gap, this paper proposes a robustness analysis based on the PI framework described in \cite{shahedi2023lead} to evaluate the impact on the system performance of how the PI-containers are divided into modules, and how long are processed at the PI-hubs. 
To do so, the variability of the system performance (represented by four KPIs) caused by a variation of the PI parameters is first assessed. Secondly, a GSA is performed to assess the individual effect of each PI parameter on the resulting system performance, allowing to evaluate the model robustness with respect to such PI parameters.

Results provide insights on which parameters should be estimated with higher accuracy or directly optimized by the optimization model.

\section{The considered optimization-based Physical Internet network} \label{problem}

This section describes the PI network and the relevant optimization problem, considered in this paper. 

\subsection{The considered PI network scheme}\label{scheme}

\begin{figure*}[t]
\centerline{\includegraphics[scale=1.5]{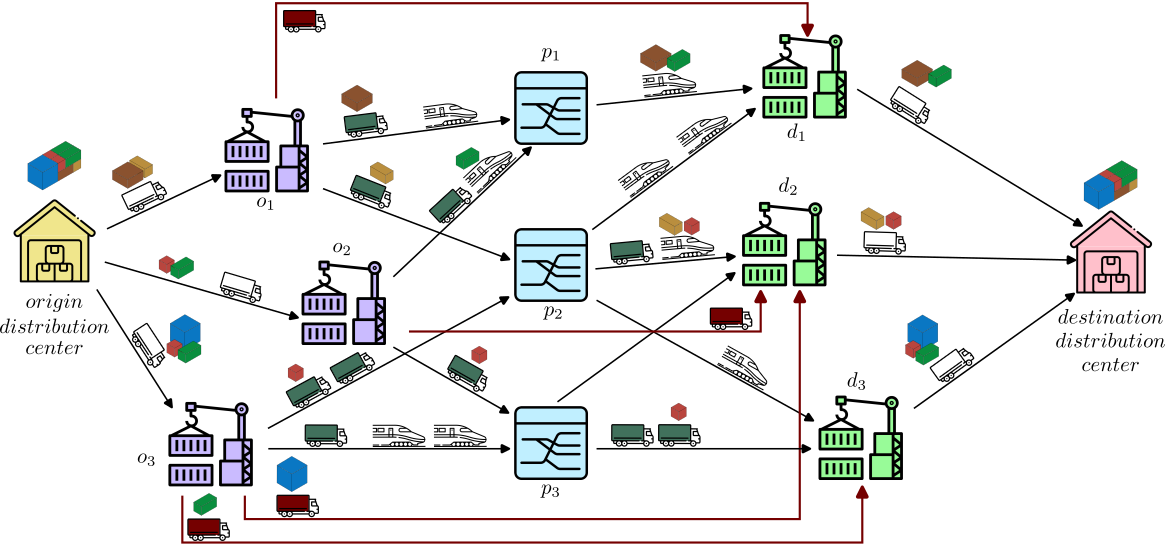}}
\caption{A diagram depicting a generic PI network, along with an illustrative example of a PI-container breakdown and routing with trucks (in green), trains, and direct trucks from origin to destination terminals (in red).}
\label{fig:network}
\end{figure*}

Let us consider the Physical Internet scheme represented in Fig.~\ref{fig:network}. This scheme, which represents a schematic example of the PI structure considered in this paper, consists of a graph of nodes representing distribution centers, terminals, and PI-hubs, and links representing the connections between these nodes.

In such a logistic network, it is assumed that PI-containers, stored in PI-containers, are initially transferred from the origin distribution centers to an origin terminal, where they can be disaggregated into smaller modules and shipped, based on the available transportation services, to the destination terminals. In this connection, such modules can be transferred either by trains and trucks through the PI-hubs network or by means of direct connections operated by trucks. Upon delivery of the last module, PI-containers are aggregated at a destination terminal and then delivered to the destination distribution center.  

In more detail, \textbf{distribution centers} serve as the starting and ending points for the PI-containers and are connected directly with suppliers and customers. The \textbf{terminals} are linked directly to both distribution centers and PI-hubs and play the most important role within the network. At these nodes, PI-containers are either disassembled into modules or assembled into a single PI-container to be delivered to the customers. Each of these nodes can be categorized as either origin or destination terminals for every PI-container. However, in general, terminals have the capability to function as both origin and destination for different PI-containers.
Finally, \textbf{PI-hubs} represent the nodes where modules can be transferred between different transportation means (e.g., from train to truck, from one truck to another, and so on). These hubs are exclusively designated for PI and enable the routing optimization in the PI logistic scheme. It is assumed that they operate rapid switches among different arriving/departing trucks and trains with a fully automated setting, without requiring extensive storage capacity.

The rail and road links in Fig.~\ref{fig:network} are operated by planned and capacitated trains and trucks, respectively, with a specific departure time for each vehicle. 

As regards operations, the considered PI network is assumed to operate according to the solution of a centralized optimization problem that provides the optimal route and specific train or truck for each module.

In addition, the following assumptions hold in the rest of the paper:
\begin{itemize}
    \item The travel times of each transportation mode on each link are fixed and a priori known.
    \item Holding modules at the destination terminals due to the arrival time gaps between the first and last module incurs inventory costs. Such costs may be assumed to be proportional to the module size and the arrival gaps. For the sake of simplicity, such costs are included in the model with a simplified approach that minimizes the gap between the arrival time of the first and last module.
\end{itemize}

\subsection{PI centralized optimization model} \label{model}

In this section, the centralized optimization problem that determines the routing and assignment of the modules is presented. 
In doing so, it is worth mentioning that the presented model was already proposed by the authors \cite{shahedi2023lead}, with small differences that will be highlighted below, and it is here summarized for the sake of clarity. The detailed notation is shown in Tables \ref{Tab.sets}, \ref{Tab.par}, and \ref{Tab.Var}.

\begin{longtable}[t]{p{1.3cm} p{15cm}}
\caption{Sets} \\
\toprule
Symbol & Meaning \\
\midrule
$\mathcal{S}$ & Set of PI-containers \\
$\mathcal{O}_{s} $ & Set of origin terminals of the PI-container $s\in\mathcal{S}$ \\
$\mathcal{D}_{s} $ & Set of destination terminals of the PI-container $s\in\mathcal{S}$ \\
$\mathcal{P} $ & Set of the PI-hubs \\
$\mathcal{N} $ & Set of all nodes, i.e., $\mathcal{N}=\mathcal{O}_{s} \cup \mathcal{D}_{s} \cup \mathcal{P}$ \\
$\mathcal{L}$ & Set of all links $(j,k)$, $\forall j,k\in\mathcal{N}$ \\
$\mathcal{J}_{s} $ & Set of the modules corresponding to the generic PI-container $s\in\mathcal{S}$ \\
$\mathcal{B}^j $ & Set gathering all the nodes $k\in\mathcal{N}$ such that a link $(k, j)$ exists \\ 
$\mathcal{F}^j $ & Set gathering all the nodes $k\in\mathcal{N}$ such that the link $(j, k)$ exists\\
$\mathcal{R}^{jk} $ & Set of trains operating on the link $(j,k)\in \mathcal{L}$ \\
$\mathcal{C}^{jk} $ & Set of trucks operating on the link $(j,k)\in \mathcal{L}$ \\
$\mathcal{T}^{jk} $ & Set of direct trucks connecting two generic nodes $j\in\mathcal{O}_{s}$ and $k\in \mathcal{D}_{s}$ \\
$\mathcal{G}^{jk}$ & Set of all the vehicles operating on the link $(j,k)\in \mathcal{L}$, i.e., 
${\mathcal{G}^{jk}}=\mathcal{R}^{jk} \cup \mathcal{C}^{jk} \cup \mathcal{T}^{jk}$ \\ 
\bottomrule
\label{Tab.sets}
\end{longtable}

\begin{longtable}[t]{p{1.3cm} p{15cm}}
\caption{Parameters} \\
\toprule
Symbol & Meaning \\
\midrule
$n_s $ & Number of modules in which a PI-container $s\in\mathcal{S}$ is divided\\
$V_{i s} $ & Volume of module $i\in\mathcal{J}_{s}$ in PI-container $s\in\mathcal{S}$ \\
$c^{j k \ell}$ & Unitary cost per distance and module for the link $(j,k)$ and transportation mode $\ell \in \mathcal{G}^{jk}$ \\
$\Gamma^{j k \ell}$ & Capacity of a generic vehicle $\ell \in \mathcal{G}^{jk}$ operating on the link $(j,k)$\\
$t^{j k \ell}$ & Travel time of the vehicle $\ell \in \mathcal{G}^{jk}$ operating on the link $(j,k)$ \\
$t^k$ & Travel time from the terminal $k \in \mathcal{D}$ to the destination distribution terminal \\
$\tau_p $ & Average operation time of the PI-hub (unloading, sorting, and reloading time for each module) $p \in \mathcal{P}$ \\
${dp}^{j k \ell} $ & Departure time of the vehicle $\ell \in \mathcal{G}^{jk}$ connecting the nodes $j \in \mathcal{N}$ and $k \in \mathcal{N}$\\
${\theta}_{s} $ & Priority associated with PI-container $s\in\mathcal{S}$ \\
$a_{i s}^k $ & Arrival time of module $i\in\mathcal{J}_{s}$ of PI-container $s$ at node $k \in \mathcal{O}_{s}$\\
$d^{j k \ell}$ & Length of the link $(j,k)\in\mathcal{L}$ when the freight is transferred by vehicle $\ell \in \mathcal{G}^{jk}$
\\
\bottomrule
\label{Tab.par}
\end{longtable}

\begin{longtable}[t]{p{1.3cm} p{15cm}}
\caption{Variables} \\
\toprule
\multicolumn{2}{c}{\textbf{Decision Variables}} \\
\midrule
Symbol & Meaning \\
\midrule
${z}_{i s}^{j k {\ell}} \quad $ & Binary variable equal to 1 if module $i$ of the PI-container $s$ is assigned to direct truck $\ell \in \mathcal{T}^{jk}$ from origin terminal $j\in\mathcal{O}_s$ to destination terminal $k\in\mathcal{D}_s$, and 0 otherwise, $\forall i\in\mathcal{I}_s, s\in\mathcal{S}$\\
${y}_{i s}^{j k {\ell}} \quad $ & Binary variable equal to 1 if module $i$ of the PI-container $s$ is assigned to train $\ell \in \mathcal{R}^{jk}$ connecting the nodes $j \in \mathcal{N}$ and $k \in \mathcal{N}$, and 0 otherwise, $\forall i\in\mathcal{I}_s, s\in\mathcal{S}$\\
${x}_{i s}^{{j k \ell}} $ & Binary variable equal to 1 if module $i$ of the PI-container $s$ is assigned to truck $\ell \in \mathcal{C}^{jk}$ connecting the nodes $j \in \mathcal{N}$ and $k \in \mathcal{N}$, and 0 otherwise, $\forall i\in\mathcal{I}_s, s\in\mathcal{S}$\\
\midrule
\multicolumn{2}{c}{\textbf{Other Variables}} \\
\midrule
Symbol & Meaning \\
\midrule
${D T}_{s}$ & Delivery time of PI-container $s \in \mathcal{S}$ \\
${D T}_{s}^{k} $ & Delivery time of PI-container $s \in \mathcal{S}$ to destination terminal $k \in \mathcal{D}_{s}$\\
${\rho}_{{i s}}^{{k}} $ & Delivery time of module $i\in\mathcal{I}_s$ of PI-container $s \in \mathcal{S}$ at the node $k \in \mathcal{P} \cup \mathcal{D}_{s}$\\
${\varphi}_{i s}^{k} $ & Delivery time of module $i\in\mathcal{I}_s$ of PI-container $s \in \mathcal{S}$ sent entirely by truck to destination terminal $k \in \mathcal{D}_{s}$\\
$\alpha_{s}$ & Delivery time of the first module of PI-container $s \in \mathcal{S}$ delivered to the destination terminal \\
$\omega_{s}$ & Delivery time of the last module of PI-container $s \in \mathcal{S}$ delivered to the destination terminal \\
\bottomrule
\label{Tab.Var}
\end{longtable}

In this paper, the mathematical programming model determining the optimal assignment and routing of modules is formulated as minimization of a single KPI 
\begin{equation}
\min J_i, \quad i=1,2,3,4
\label{eq:CF}
\end{equation}
where
\begin{equation}
J_{1}=\sum_{s \in \mathcal{S}} \sum_{i \in \mathcal{J}_{S}} \sum_{j \in \mathcal{O}_{s}} \sum_{k \in \mathcal{D}_{s}} \sum_{\ell \in \mathcal{T}^{jk}} z_{i s}^{j k \ell} \label{1}
\end{equation}
represents the total usage of the direct trucks for the shipping operations between terminals,
\begin{equation}
J_{2}=\sum_{k \in \mathcal{D}_{S}} \sum_{s \in \mathcal{S}} \theta_{s} \cdot D T_{s}^{k} \label{2}
\end{equation}
represents the total delivery time for each PI-container,
\begin{multline}
J_{3} = \sum_{i \in \mathcal{J}_{s}} \sum_{s \in \mathcal{S}}\left(\sum_{j, k \in \mathcal{N}}\left(\sum_{\ell \in \mathcal{R}^{jk}} c^{j k \ell} \cdot V_{is} \cdot d^{j k \ell} \cdot y_{i s}^{j k \ell}+\sum_{\ell \in \mathcal{C}^{jk}} c^{j k \ell} \cdot V_{is} \cdot d^{j k \ell} \cdot x_{i s}^{j k \ell}\right)\right. + \\
\left. + \sum_{j \in \mathcal{O}_{s}} \sum_{k \in \mathcal{D}_{s}} \sum_{\ell \in \mathcal{T}^{jk}} c^{j k \ell} \cdot V_{is} \cdot d^{j k \ell} \cdot z_{i s}^{j k \ell} \right)
\label{3}
\end{multline}
represents the total cost of the transportation of the PI-container in the network, and
\begin{equation}
J_{4}=\sum_{s \in \mathcal{S}}  \omega_s - \alpha_s \label{4}
\end{equation}
represents the total temporal difference between the delivery times of the last arrived module of PI-containers ($\omega_s$) and the delivery times of the first one ($\alpha_s$).

Note that the cost function in Eq.\eqref{eq:CF} represents a significant difference compared to (\cite{shahedi2023lead}), where all the KPIs $J_i$ were considered in a weighted sum. The aim of this choice, which will be further clarified in the robustness analysis, is to separate the effects of the different parameters on each KPI. Furthermore, unlike (\cite{shahedi2023lead}), parametric transportation costs (i.e., costs per traveled distance and volume of the PI-container) and an additional KPI ($J_4$) are considered here.

As regards the constraints, they are formulated as
\begin{gather}
DT_s^k \geq \varphi_{i s}^{k}, \quad  \forall i \in \mathcal{J}_{s}, s \in \mathcal{S}, k \in \mathcal{D}_{s} \label{5} \\
DT_s^k \geq \rho_{i s}^{k}, \quad \forall i \in \mathcal{J}_{s}, s \in \mathcal{S}, k \in \mathcal{D}_{s} \label{6} 
\end{gather}
defining the delivery time of each PI-container at the destination terminal as the delivery time of the latest of its modules;
\begin{gather}
DT_s \geq \{DT_{s}^{k} + t^k\},\quad \forall s \in \mathcal{S}, \forall k\in \mathcal{D}_{s} 
\label{7} 
\end{gather}
which define the delivery time of each PI-container at the destination distribution center as the maximum delivery time at the destination terminal, where the modules are reassembled, plus the travel time required to reach the destination distribution center;
\begin{gather}
a_{is}^{j} \leq dp^{jk\ell} x_{is}^{jk\ell} + M\left(1-x_{i s}^{j k \ell}\right),\quad \forall \ell \in \mathcal{C}^{jk}, j \in \mathcal{O}_{s}, k \in \mathcal{P}, i \in \mathcal{J}_{s}, s \in \mathcal{S} \label{8} \\
a_{i s}^{j} \leq dp^{j k \ell} y_{i s}^{j k \ell}+M\left(1-y_{i s}^{j k \ell}\right),\quad \forall \ell \in \mathcal{R}^{jk}, j \in \mathcal{O}_{s}, k \in \mathcal{P}, i \in \mathcal{J}_{s}, s \in \mathcal{S} \label{9} \\
a_{i s}^{j} \leq dp^{j k \ell} z_{i s}^{j k \ell}+M\left(1-z_{i s}^{j k \ell}\right), \quad  \forall \ell \in \mathcal{T}^{jk}, j \in \mathcal{O}_{s}, k \in \mathcal{D}_{s}, i \in \mathcal{J}_{s}, s \in \mathcal{S} \label{10} 
\end{gather}
which state that the departure time of each module from an origin terminal must be greater than or equal to its arrival time at that terminal;
\begin{gather}
\rho_{i s}^{k} \geq dp^{j k \ell}+t^{j k \ell}-M\left(1-x_{i s}^{j k \ell}\right) , \quad \forall \ell \in \mathcal{C}^{jk}, k \in \mathcal{P} \cup \mathcal{D}_{s}, j \in \mathcal{B}_{k}, i \in \mathcal{J}_{s}, s \in \mathcal{S} \label{11} \\
\rho_{i s}^{k} \geq dp^{j k \ell}+t^{j k \ell}-M\left(1-y_{i s}^{j k \ell}\right) , \quad \forall \ell \in \mathcal{R}^{jk}, k \in \mathcal{P} \cup \mathcal{D}_{s}, j \in \mathcal{B}_{k}, i \in \mathcal{J}_{s}, s \in \mathcal{S} \label{12} 
\end{gather}
which define the arrival time at the nodes for each module transferred from the origin terminals to the PI-hubs by trucks and trains, respectively;
\begin{gather}
\rho_{i s}^{j} \leq dp^{j k \ell} x_{i s}^{j k \ell}+\tau_p+M\left(1-x_{i s}^{j k \ell}\right) , \quad \forall \ell \in \mathcal{C}^{jk}, j \in \mathcal{P}, k \in \mathcal{F}_{k}, i \in \mathcal{J}_{s}, s \in \mathcal{S} \label{13} \\
\rho_{i s}^{j} \leq dp^{j k \ell} y_{i s}^{j k \ell}+\tau_p+M\left(1-y_{i s}^{j k \ell}\right), \quad \forall \ell \in \mathcal{R}_{k, h}, j \in \mathcal{P}, k \in \mathcal{F}_{k}, i \in \mathcal{J}_{s}, s \in \mathcal{S} \label{14} 
\end{gather}
which state that modules cannot depart from PI-hubs towards the destination terminals by trucks or trains before arriving;
\begin{gather}
\varphi_{i s}^{k} \geq dp^{j k \ell}+t^{j k \ell}-M\left(1-z_{i s}^{j k \ell}\right), \quad \forall \ell \in \mathcal{T}^{jk}, j \in \mathcal{O}_{s}, k \in \mathcal{D}_{s}, i \in \mathcal{J}_{s}, s \in \mathcal{S} \label{15} 
\end{gather}
which defines the delivery time of each module at the destination terminal transferred by means of direct trucks;
\begin{gather}
\sum_{s \in \mathcal{S}} \sum_{i \in \mathcal{J}_{S}} V_{i s} x_{i s}^{j k \ell} \leq \Gamma^{j k \ell}, \quad \forall \ell \in \mathcal{C}^{jk}, \forall j, k \in \mathcal{N}, i \in \mathcal{J}_{s}, s \in \mathcal{S} \label{16} \\
\sum_{s \in \mathcal{S}} \sum_{i \in \mathcal{J}_{s}} V_{i s} y_{i s}^{j k \ell} \leq \Gamma^{j k \ell}, \quad \forall \ell \in \mathcal{R}^{jk}, \forall j, k \in \mathcal{N}, i \in \mathcal{J}_{s}, s \in \mathcal{S} \label{17} \\
\sum_{s \in \mathcal{S}} \sum_{i \in \mathcal{J}_{s}} V_{i s} z_{i s}^{j k \ell} \leq \Gamma^{j k \ell}, \quad \forall \ell \in \mathcal{T}^{jk}, j \in \mathcal{O}_{s}, k \in \mathcal{D}_{s}, i \in \mathcal{J}_{s}, s \in \mathcal{S} \label{18} 
\end{gather}
which guarantee that the modules assigned to each transportation mode do exceed its capacity;
\begin{equation}
\sum_{o \in \mathcal{O}_s}\sum_{k \in \mathcal{F}_{o}}\left(\sum_{\ell \in \mathcal{R}_{o, k}} x_{i s}^{o k \ell}+\sum_{\ell \in \mathcal{C}_{o, k}} y_{i s}^{o k \ell} +\sum_{\ell \in \mathcal{T}_{o, k}} z_{i s}^{o k \ell} \right)=1, \quad \forall i \in \mathcal{J}_{s}, \forall s \in \mathcal{S}\label{19}
\end{equation}
\begin{equation}
\sum_{d \in \mathcal{D}_s}\sum_{k \in \mathcal{B}_{d}}\left(\sum_{\ell \in \mathcal{R}_{k, d}} x_{i s}^{k d \ell}+\sum_{\ell \in \mathcal{C}_{k, d}} y_{i s}^{k d \ell} +\sum_{\ell \in \mathcal{T}_{k, d}} z_{i s}^{k d \ell} \right)=1, \quad \forall i \in \mathcal{J}_{s}, \forall s \in \mathcal{S}\label{20}
\end{equation}
which guarantee that each PI-container starts (resp., ends) its trip from an origin (resp., at a destination) terminal by means of a train, a truck directed to a PI-hub, or a direct truck directed to a destination terminal. For each module, the variable $x_{i s}^{o k \ell}$, $y_{i s}^{o k \ell}$, or $z_{i s}^{o k \ell}$ set to 1 in the optimal solution indicates, for the module $i$ of the PI-container $s$, the assigned origin terminal $o$. Analogously, the variable $x_{i s}^{dk\ell}$, $y_{i s}^{dk\ell}$, or $z_{i s}^{dk\ell}$ set to 1 in the optimal solution indicates the assigned destination terminal $d$;
\begin{gather}
\sum_{j \in \mathcal{O}_{s}}\left(\sum_{\ell \in \mathcal{R}_{j, p}} x_{i s}^{j p \ell}+\sum_{\ell \in \mathcal{C}_{j, p}} y_{i s}^{j p \ell}\right)=\sum_{k \in \mathcal{D}_{s}}\left(\sum_{\ell \in \mathcal{R}_{p, k}} x_{i s}^{p k \ell}+\sum_{\ell \in \mathcal{C}_{p, k}} y_{i s}^{p k \ell}\right), \quad \forall i \in \mathcal{J}_{s}, \forall s \in \mathcal{S}, p \in \mathcal{P} \label{23} 
\end{gather}
which guarantees that any PI-container passes through only one PI-hub, unless it is delivered by direct truck;
\begin{multline}
\alpha_s \leq \sum_{{j,k} \in \mathcal{N}} \left(\sum_{\ell \in \mathcal{R}^{jk}} \left(d^{j k \ell} + t^{j k \ell}\right) \cdot y_{i s}^{j k \ell} + \sum_{\ell \in \mathcal{C}^{jk}} \left(d^{j k \ell} + t^{j k \ell}\right) \cdot x_{i s}^{j k \ell} \right) + \\
+ \sum_{j \in \mathcal{O}_{s}} \left(\sum_{k \in \mathcal{D}_{s}} \sum_{\ell \in \mathcal{T}^{jk}} t^{j k \ell} \cdot z_{i s}^{j k \ell} 
\sum_{k \in \mathcal{P}} \sum_{\ell \in \mathcal{R}^{jk} \cup \mathcal{C}^{jk}} \tau_{p} \cdot (y_{i s}^{j k \ell} + x_{i s}^{j k \ell})\right), \quad \forall i \in \mathcal{J}_{s}, \forall s \in \mathcal{S} \label{24}
\end{multline}
\begin{multline}
\omega_s \geq \sum_{{j,k} \in \mathcal{N}} \left(\sum_{\ell \in \mathcal{R}^{jk}} \left(d^{j k \ell} + t^{j k \ell}\right) \cdot y_{i s}^{j k \ell} +\sum_{\ell \in \mathcal{C}^{jk}} \left(d^{j k \ell} + t^{j k \ell}\right) \cdot x_{i s}^{j k \ell} \right) + \\
+ \sum_{j \in \mathcal{O}_{s}} \left( \sum_{k \in \mathcal{D}_{s}} \sum_{\ell \in \mathcal{T}^{jk}} t^{j k \ell} \cdot z_{i s}^{j k \ell} 
+ \sum_{k \in \mathcal{P}} \sum_{\ell \in \mathcal{R}^{jk} \cup \mathcal{C}^{jk}} \tau_{p} \cdot (y_{i s}^{j k \ell} + x_{i s}^{j k \ell})\right), \quad \forall i \in \mathcal{J}_{s}, \forall s \in \mathcal{S} \label{25}
\end{multline}
which define, respectively, the delivery time of the first and last module of a PI-container. 
Constraints in Eq.~\eqref{24} and \eqref{25} are introduced for the first time in this paper and are used in Eq.~\eqref{4} to define $KPI_4$.

\section{Robustness analysis}
\label{sec:gsa}

In the model described in Section~\ref{model}, the values of the operational times at the PI-hubs ($\tau_p$) and the number of modules ($n_s$) in which a PI-container is divided are assumed known in advance. However, such a condition do not always hold in real cases, where these values may be affected by uncertainty. Therefore, it is important to assess the variability introduced by the parameter uncertainty within the considered PI framework.

To this aim, a robustness analysis can be performed. In this framework, the goal will consist of the identification, among all the uncertain input parameters of a model, of the most important ones in determining the uncertainty in the output of interest.

To achieve that, in this work a variance-based GSA has been conducted; its main theoretical concepts are summarized in the following section. For further details on this topic, the reader can refer to (\cite{sal:gsa}).

\subsection{Variance-based global sensitivity analysis}
\label{Sec:sensan}

Variance-based GSA aims to decomposing the variance of a considered model output into fractions that can be attributed to the uncertain input parameters that may vary. 

From a formal point of view, the input-output dependence can be expressed as a scalar function $f: \mathbb{R}^n \rightarrow \mathbb{R}$ that produces a scalar random output $Y$ given a vector of $n$ random inputs $X = \left\{X_1, X_2, ... X_n\right\}$. 

Then, let $V_Y$ be the total variance of the output $Y$ when all the random inputs $X$ are allowed to vary, and let $V_{\sim X_i}(Y|X_i)$ be the variance when all the random inputs in $X$ except for $X_i$, are allowed to vary. With these definitions, ${\rm E}_{X_i}(V_{\sim X_i}(Y|X_i))$ represents the expectation of residual variance $V_{\sim X_i}(Y|X_i)$ ``weighed'' by the distribution of $X_i$ and can be considered as a measure of how influential the variable $X_i$ is on $Y$. Specifically, if ${\rm E}_{X_i}(V_{\sim X_i}(Y|X_i))$ is small, meaning that the variance of $Y$ is small when $X_i$ is fixed, the variability of $X_i$ has a significant influence on $V_Y$. In fact, since
\begin{equation}
V_Y = {\rm E}_{X_i}(V_{\sim X_i}(Y|X_i)) + V_{\sim X_i}({\rm E}_{X_i}(Y|X_i)),
\label{eq:var}           
\end{equation}
if $X_i$ is greatly influential, then $V_{\sim X_i}({\rm E}_{X_i}(Y|X_i))$ tends to coincide with the total variance $V_Y$. Thus, it is possible to define the sensitivity index
\begin{equation}
S_i = \frac{V_{\sim X_i}({\rm E}_{X_i}(Y|X_i))}{V_Y},
\label{eq:sobindex}           
\end{equation}
which is known in the literature as \textit{Sobol first-order index}, \textit{main effect}, or also \textit{importance measure}. Such an index represents the expected reduction in the variance of the output that could be obtained if an individual factor $X_i$ were fixed.
$S_i$ is scaled in the range $[0,1]$ and is closer and closer to $1$ the more influential $X_i$ is on $Y$.

Note that, given its definition, $S_i$ describes only the \textit{individual} effect of a single input on the output variance. However, it may happen that a given input is relevant on the output (e.g., it determines extreme output values) only when it is combined to a specific value of another input. These combined effects,  called \textit{interactions}, describe the synergistic effects that could be associated to particular combinations of two or more model inputs. Interactions become relevant if all the first order indices are very small, and so they are not able to explain the majority of the output variance. Such interactions can be evaluated by finding the so-called \textit{total-effect} terms which, however, are more complex and computationally hard to find.

It is important to note that different values of the indices must be calculated separately for any different output. Additionally, the values of the first-order index are typically  estimated numerically by generating an appropriate number of input parameters samples using Monte Carlo methods and approximated formulas. However, these methods require to evaluate the function $f$ (i.e., to run the optimization model) multiple times, which becomes a limitation when the model takes a significant amount of time for a single run.
Additionally, if more than one KPI is considered as output, the number of required executions increases proportionally to the number of outputs considered. To address this issue, in this paper the values of the first-order index are estimated using a more advanced method that applies the same basic GSA's idea by combining Random Balance Design with the Fourier Amplitude Sensitivity Test, as described in (\cite{tar:gsa}). Moreover, this method has been adapted to compute the values of the sensitivity indices without having to run the model a proportional number of times. Consequently, the number of required model runs is always equal to $N$, where $N$ is the number of samples of the inputs generated, independently from the number of outputs considered in the analysis.

\section{Case study and results} 

The robustness analysis is applied to a case study, described in Section~\ref{case_study}, which includes both realistic and assumed data. The results of the optimization model for the deterministic case, i.e., without inputs and parameters variability, are presented in Section~\ref{reference}. Finally, Section~\ref{robustness_results} presents the results of the robustness analysis. 

\subsection{Case study description} 
\label{case_study}

\begin{figure*}[h]
\centerline{\includegraphics[scale=0.50]{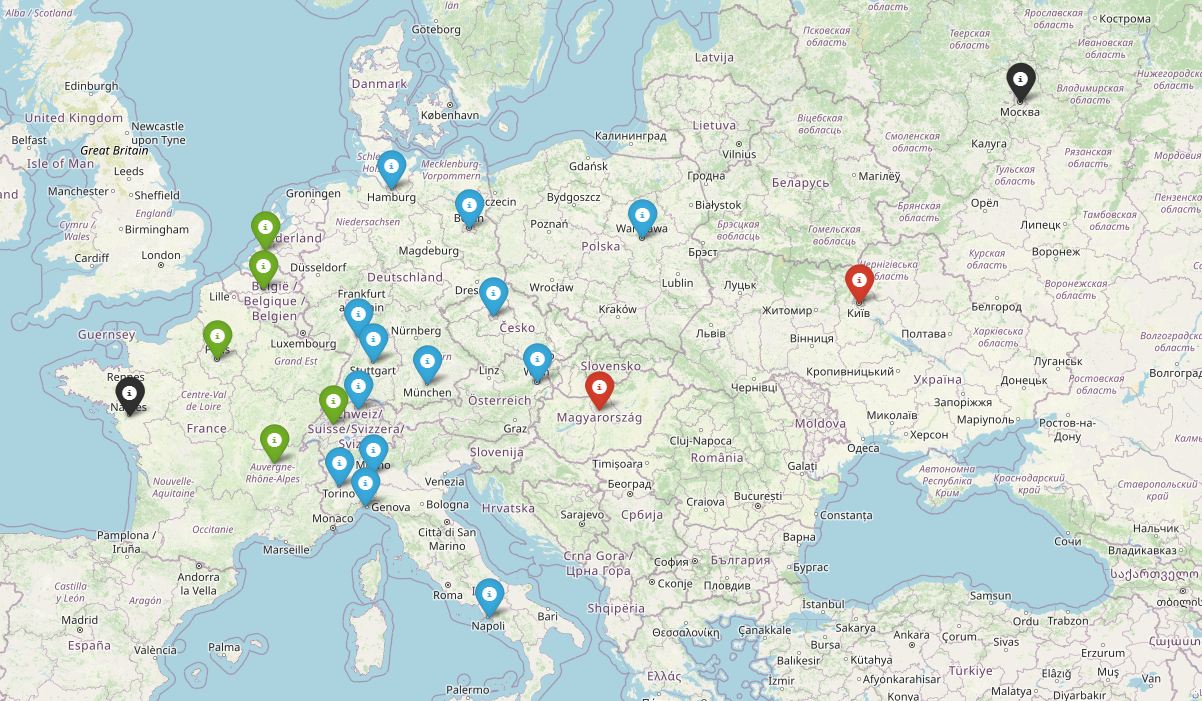}}
\caption{The assumed location of the terminals and PI-hubs in Europe - Black spots: distribution centers; Green spots: origin terminals; Blue spots: PI-hubs; Red spots: destination terminals.}
\label{map}
\end{figure*}

The considered case study consists of a PI logistic network with $|\mathcal{O}|=5$ origin terminals, $|\mathcal{P}|=13$ PI-hubs, $|\mathcal{D}|=2$ destination terminals, along with one origin and one destination distribution center, as depicted in Fig.~\ref{map}. The origin and destination distribution centers are located in Nantes and Moscow, respectively. The origin terminals (in green) are situated in Rotterdam, Paris, Lyon, Brussels and Bern. The PI-hubs are located in Hamburg, Mannheim, Berlin, Munich, Stuttgart, Milan, Turin, Genoa, Naples, Prague, Zurich, Vienna, and Warsaw. Additionally, there are 2 destination terminals in Budapest, and Kyiv.

Regarding the freights to be delivered, $|\mathcal{S}|=20$ PI-container are considered, each consisting of a fix number of modules in the range $[1,10]$. Then, concerning the transportation modes available in the considered network, the PI-containers can be transferred from the origin terminals to the destination terminals either by direct trucks or via the PI-hubs using combination of trains and trucks. In this context, $|\mathcal{R}^{jk}|=3$ trains and $|\mathcal{C}^{jk}|=15$ trucks are considered for each link $(j,k)\in\mathcal{L}$, while $|\mathcal{T}^{jk}|=20$ direct trucks are available to directly travel between the origin and destination terminals. The departure times and available capacities for all transportation modes and are known a priori and are reported in Tables \ref{tab:train-info}, \ref{Table 3}, and \ref{Table 4}. 

The price for shipping one container by truck or train in Europe varies depending on several factors, including country, fuel costs, container's size, etc. Based on scholarly references, like (\cite{KIM2011859}), the transportation cost of one 20ft PI-container is estimated at 1.2 \euro/kilometer and 0.62 \euro/kilometer for trucks and trains respectively. Additionally, this price is estimated at 3.78 \euro/kilometer for direct trucks.

Finally, the operation times required at the PI-hubs to transfer the modules from the incoming vehicles to the outgoing ones are reported in Table ~\ref{Tab.OP.time}.

\begin{table}[t]
\centering
\caption{Trains Information}
\begin{tabular}{c|ccc|ccc}
\hline
\textbf{} & \multicolumn{3}{|c|}{Train in link $(j,k)$, $j \in \mathcal{O}_{s}$ , $k \in \mathcal{P}$} & \multicolumn{3}{|c}{Train in link $(j,k)$, $j \in \mathcal{P}$, $k \in \mathcal{D}_{s}$} \\
\hline
Train & 1 & 2 & 3 & 1 & 2 & 3 \\
\hline
Departure time $({dp}^{{j k} {\ell}})$ & 9:00 & 13:00 & 20:00 & 9:00 & 13:00 & 20:00 \\
Capacity $(\Gamma^{j k \ell})$ in $m^3$ & 21 & 18 & 26 & 25 & 21 & 27 \\
\hline
\end{tabular}
\label{tab:train-info}
\end{table}

\begin{table}[htbp]
\centering
\caption{Information of the trucks in each link (between terminals and PI-hubs) which can commute daily}
\resizebox{\textwidth}{!}{
\begin{tabular}{cccccccccccccccc}
\toprule
Truck $\ell \in \mathcal{C}^{jk}$ in link \\
$(j,k)$, $j \in \mathcal{O}_{s}$ , $k \in \mathcal{P}$ & 1 & 2 & 3 & 4 & 5 & 6 & 7 & 8 & 9 & 10 & 11 & 12 & 13 & 14 & 15 \\
\midrule
Capacity $(\Gamma^{j k \ell})$ in $m^3$ & 8 & 9 & 6 & 8 & 7 & 8 & 11 & 10 & 12 & 10 & 9 & 10 & 6 & 15 & 11 \\
\cline{1-1}
Departure time  $({dp}^{{j k} {\ell}})$ & 6:30 & 7:30 & 8:30 & 9:30 & 11:30 & 12:30 & 13:30 & 14:30 & 16:00 & 17:30 & 18:30 & 19:30 & 20:30 & 21:30 & 23:30 \\
\midrule
Truck $\ell \in \mathcal{C}^{jk}$ in link \\
$(j,k)$, $j \in \mathcal{P}$, $k \in \mathcal{D}_{s}$ & 1 & 2 & 3 & 4 & 5 & 6 & 7 & 8 & 9 & 10 & 11 & 12 & 13 & 14 & 15 \\
\midrule
Capacity $(\Gamma^{j k \ell})$ in $m^3$  & 8 & 7 & 5 & 9 & 6 & 8 & 8 & 7 & 9 & 7 & 8 & 10 & 11 & 12 & 10 \\
\cline{1-1}
Departure time  $({dp}^{{j k} {\ell}})$ & 6:30 & 7:30 & 8:30 & 9:30 & 11:30 & 12:30 & 13:30 & 14:30 & 16:00 & 17:30 & 18:30 & 19:30 & 20:30 & 21:30 & 23:30 \\
\bottomrule
\end{tabular}}
\label{Table 3}
\end{table}

\begin{table}[htbp]
\centering
\caption{Capacity of the direct trucks in each link (between terminals) which can commute daily $(\Gamma^{j k \ell})$}
\resizebox{\textwidth}{!}{
\begin{tabular}{cccccccccccccccc}
\toprule
Direct Truck $\ell \in \mathcal{T}^{jk}$ & 1 & 2 & 3 & 4 & 5 & 6 & 7 & 8 & 9 & 10 \\
\cline{1-1}
Capacity $(\Gamma^{j k \ell})$ of direct truck in link $(j,k)$, $j \in \mathcal{O}_{s}, k \in \mathcal{D}_{s}$ & 10 & 9 & 6 & 12 & 7 & 8 & 6 & 8 & 4 & 5 \\
\midrule
Direct Truck $\ell \in \mathcal{T}^{jk}$ & 11 & 12 & 13 & 14 & 15 & 16 & 17 & 18 & 19 & 20 \\
\cline{1-1}
Capacity $(\Gamma^{j k \ell})$ of direct truck in link $(j,k)$, $j \in \mathcal{O}_{s}, k \in \mathcal{D}_{s}$ & 10 & 9 & 6 & 12 & 7 & 8 & 6 & 2 & 4 & 5 \\
\bottomrule
\end{tabular}}
\label{Table 4}
\end{table}

\begin{table}[h]
\centering
\caption{The average operation time $(\tau_p)$ of each PI-hub for transferring the modules from the incoming section to the outgoing one}
\resizebox{\textwidth}{!}{
\begin{tabular}{cccccccc}
\toprule
PI-hub $p \in \mathcal{P}$ & Hamburg & Mannheim  & Milan  & Berlin & Prague & Munich  & Zürich  \\
\cline{1-1}
Average operation time $(\tau_p)$ (hrs) & 2 & 1 & 1 & 2 & 2 & 1 & 2 \\
\midrule
PI-hub $p \in \mathcal{P}$ & Stuttgart & Turin & Genoa  & Naples & Vienna  & Warsaw  & \\
\cline{1-1}
Average operation time $(\tau_p)$ (hrs) & 2 & 3 & 1 & 1 & 2 & 1 & \\
\bottomrule
\end{tabular}}
\label{Tab.OP.time}
\end{table}

\subsection{Optimal results in deterministic case} \label{reference}

In order to test the optimization model and provide evidence of its effectiveness, the problem in \eqref{eq:CF}--\eqref{25} was first solved by assuming all the constant and a priori known parameters introduced in the previous section. 
The problem was solved using CPLEX$^\copyright$ on a workstation equipped with an Intel\textsuperscript{TM} Core i7-CPU running at 3.30 GHz and 16.00 GB of RAM, with an average solution time of almost 4 minutes. 

For the sake of compactness, the main characteristics of the obtained solution are reported in Table~\ref{Table 8} for the case of $J_2$ minimization, which is the most common objective in logistics optimization (\cite{sun2018multiagent, chen2020minimizing}).
In this table, the entries specify the modules of each PI-container that are transported from the origin terminals to the destination terminals using direct trucks or various combinations of vehicles (trains/trucks) through PI-hubs.

\begin{table}[htbp]
\centering
\caption{PI-container and Modules transferred through Pi-hubs and via Direct trucks. The first and last modules arrived at the destination are identified by italic red and bold, respectively.}
\resizebox{\textwidth}{!}{
\begin{tabular}{c|c|c|c|c|c|c|c|c|c|c}
\toprule
\textbf{PI-container $s$} & $s=1$ & $s=2$ & $s=3$ & $s=4$ & $s=5$ & $s=6$ & $s=7$ & $s=8$ & $s=9$ & $s=10$ \\
\textbf{Number of modules $n_s$} & $n_1=3$ & $n_2=5$ & $n_3=2$ & $n_4=6$ & $n_5=7$ & $n_6=8$ & $n_7=8$ & $n_8=9$ & $n_9=10$ & $n_{10}=5$ \\
\midrule
\textbf{Routed via Pi-hub} & \multirow{2}{*}{1, 2, 3}& \multirow{2}{*}{-} & \multirow{2}{*}{1, 2} & \multirow{2}{*}{4} & \multirow{2}{*}{2, 3} & \multirow{2}{*}{4, 6} & \multirow{2}{*}{1, 3, 5} & \multirow{2}{*}{2} & \multirow{2}{*}{1, 2, 8} & \multirow{2}{*}{2} \\
\textbf{Train/Train} & &  & & & & & & & & \\
\textbf{Train/Truck} & - & 1, 4 & - & 2, 5 & 1, 7 & 5 & - & - & - & 3 \\
\textbf{Truck/Train} & - & 2 & - & 3 & 6 & 8 & 7 & 1, 8 & 3, 5, 7 & 4, 5 \\
\textbf{Truck/Truck} & - & - & - & - & 4 & 2 & - & - & - & 1 \\
\midrule
\textbf{Routed via Direct trucks} & - & 3, 5 & - & 1, 6 & 5 & 1, 3, 7 & 2, 4, 6, 8 & 3, 4, 5, 6, 7, 9 & 4, 6, 9, 10 & - \\
\midrule
\midrule
\textbf{PI-container $s$} & $s=11$ & $s=12$ & $s=13$ & $s=14$ & $s=15$ & $s=16$ & $s=17$ & $s=18$ & $s=19$ & $s=20$ \\
\textbf{Number of modules $n_s$} & $n_1=6$ & $n_2=4$ & $n_3=3$ & $n_4=7$ & $n_5=5$ & $n_6=4$ & $n_7=6$ & $n_8=9$ & $n_9=10$ & $n_{10}=8$ \\
\midrule
\textbf{Routed via Pi-hub} & \multirow{2}{*}{2, 4}& \multirow{2}{*}{1, 2, 4} & \multirow{2}{*}{1, 2, 3} & \multirow{2}{*}{5} & \multirow{2}{*}{5} & \multirow{2}{*}{1} & \multirow{2}{*}{5, 6} & \multirow{2}{*}{1, 9} & \multirow{2}{*}{2, 3, 4, 5, 6, 8} & \multirow{2}{*}{-} \\
\textbf{Trains/Trains} &  &  &  &  &  &  &  &  &  &  \\
\textbf{Trains/Trucks} & - & 3 & - & 7 & 1, 4 & 2, 3 & - & - & 10 & 2 \\
\textbf{Trucks/Trains} & 1, 3, 6 & - & - & 6 & 3 & - & - & 5 & - & 1, 5, 7 \\
\textbf{Trucks/Trucks} & - & - & - & 1 & - & - & 4 & - & 7 & - \\
\midrule
\textbf{Routed via Direct trucks} & 5 & - & - & 2, 3, 4 & 2 & 4 & 1, 2, 3 & 2, 3, 4, 6, 7, 8 & 1, 9 & 3, 4, 6, 8 \\
\bottomrule
\end{tabular}}
\label{Table 8}
\end{table}

\begin{figure*}[h]
\centerline{\includegraphics[scale=0.51]{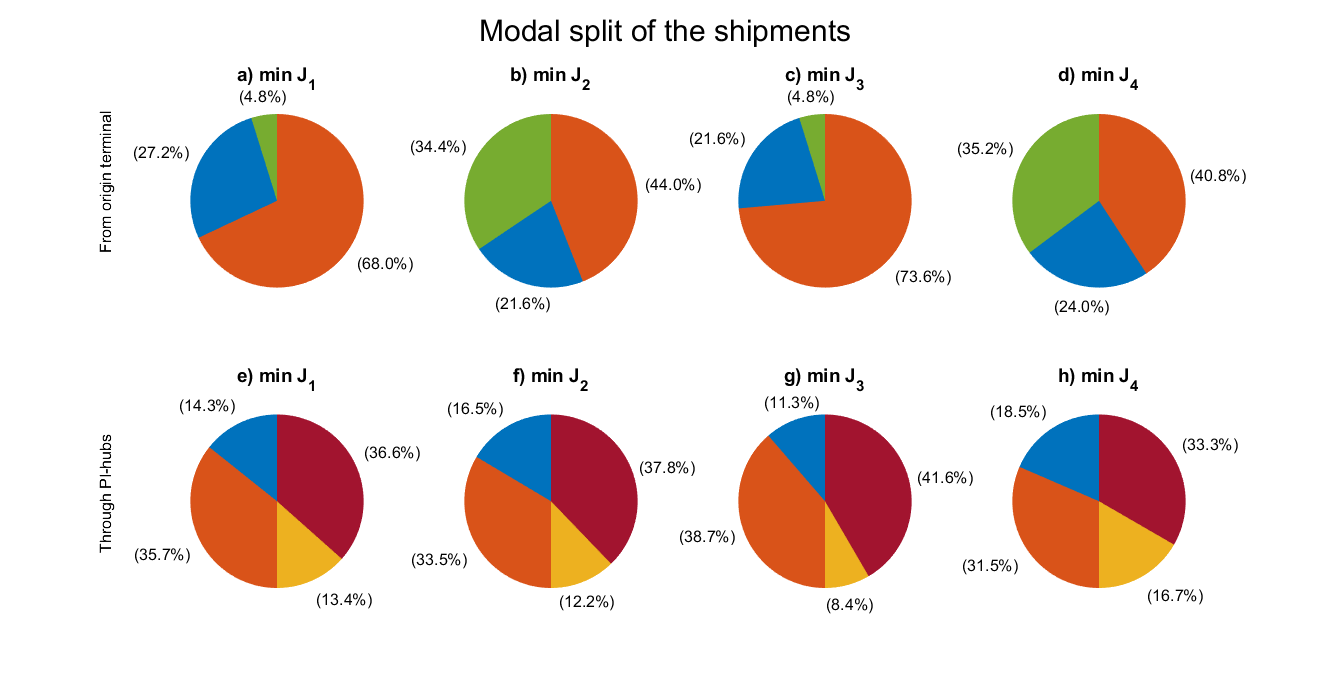}}
\caption{Modal split of the PI-containers when they leaves the origin terminal (a, b, c, d) and through the PI-hubs (e, f, g, h) for different objective functions. \textcolor[RGB]{119, 172, 48}{$\blacksquare$} Direct trucks; \textcolor[RGB]{0, 114, 189}{$\blacksquare$} Trucks to the PI-hubs; \textcolor[RGB]{217, 83, 25}{$\blacksquare$} Trains to the PI-hubs; \textcolor[RGB]{237, 177, 32}{$\blacksquare$} Trucks to destination terminals (from the PI-hubs);  \textcolor[RGB]{162, 20, 47}{$\blacksquare$} Trains to destination terminals (from the PI-hubs).}
\label{modal split}
\end{figure*}

To assess the individual effect of each objective on the obtained solutions, the optimization problem was solved by considering one of the four objectives at a time. The results show that the ratio of transportation modes varies with the different objectives as expected, thereby validating the effectiveness of the considered problem formulation. These results are presented in Fig.~\ref{modal split}, where the percentages of modules transferred by each transportation mode in the four scenarios corresponding to the optimization of a single KPI $J_i$, $i=1,2,3,4$, are shown.

In particular, the first row shows the modal split of the PI-containers when they leave the origin terminal. In such graphs, it can be noted that the direct truck usage is effectively minimized as expected when it is the main optimization goal (Fig.~\ref{modal split}-a) and also when transportation cost is the main optimization goal (Fig.~\ref{modal split}-c). Thus, these two optimization goals result into similar modal splits. On the contrary, minimizing the delivery time (Fig.~\ref{modal split}-b) or the delivery gap (Fig.~\ref{modal split}-d) leads to an increase in the use of direct trucks, as they are faster than the passage through the PI-hubs. 

The second row of Fig.~\ref{modal split} shows the modal split of the modules through the PI-hubs from the input point of view (blue and orange) and from the output point of view (yellow and red). 
In fact, trains are always the most used mode, with a modal split ranging between 31\% and 39\% at the arrival, and between 33\% and 42\% at the departure from the PI-hubs. However, similarly to the previous case, the truck usage is lower when the costs are minimized (Fig.~\ref{modal split}-g) and higher when the delivery time (Fig.~\ref{modal split}-f) or the delivery gap (Fig.~\ref{modal split}-h) are minimized.

\subsection{Robustness analysis results} \label{robustness_results}

This section presents and discusses the results of the robustness analysis conducted by on the network shown in Fig.~\ref{fig:network}, whose optimized dynamics is provided by the model in Eq.\eqref{eq:CF}--\eqref{25}. In particular, the global variability of the system performance when the input parameter varies is first investigated (Section~\ref{variability_results}). Secondly, the results of the GSA, discussing the    \textit{individual sensitivity} of each input parameter is analyzed (Section~\ref{res:gsa}).

In this analysis, the considered uncertain input parameters are the processing times $\tau_p$ at each of the PI-hubs and the number of modules in each PI-container $n_s$. In other words, this analysis evaluated how the PI logistic framework is robust with respect to how the PI-containers are divided and processed at the PI-hubs, i.e., how the system performance is affected by such aspects.
The input parameters are assumed to be stochastic independent variables with the following uniform distributions, that were chosen to represent a situation of complete uncertainty. More specifically:
\begin{itemize}
    \item $\tau_p\sim\mathcal{U}_{[1,3]}$, $\forall p\in\mathcal{P}$, i.e., the processing time for each module at the PI-hubs is modeled as random variable with uniform distribution in interval $[1,3]$ hours;
    \item $n_s\sim\mathcal{U}_{[1,10]}$, $\forall s\in\mathcal{S}$, i.e., the number of modules per PI-container is modeled as random variable with uniform distribution in interval $[1,10]$.
\end{itemize}

As for the output variables for the GSA, four KPIs are considered, each of them corresponding to one of the objective functions to minimize (i.e., ${KPI_i}:= {J_i}$, $i = 1, \ldots, 4$). In this way, each KPI represents a different performance measure of the considered PI logistic framework.

In addition, four different configurations of the problem are considered, that is:
\begin{itemize}
    \item \textbf{C1}: In the first configuration, the use of direct trucks $J_1$ is minimized;
    \item \textbf{C2}: In the second configuration, the total delivery time $J_2$  is minimized;
    \item \textbf{C3}: In the third configuration, the total transportation costs $J_3$ are minimized;
    \item \textbf{C4}: In the fourth configuration, the total delivery gap $J_4$ is minimized.
\end{itemize}

The goal of the four configurations is to test the effect of the selected objective function on the robustness of the PI framework.

\subsubsection{Analysis of the KPIs variability} \label{variability_results}

\begin{figure*}[t!]
\centerline{\includegraphics[scale=0.58]{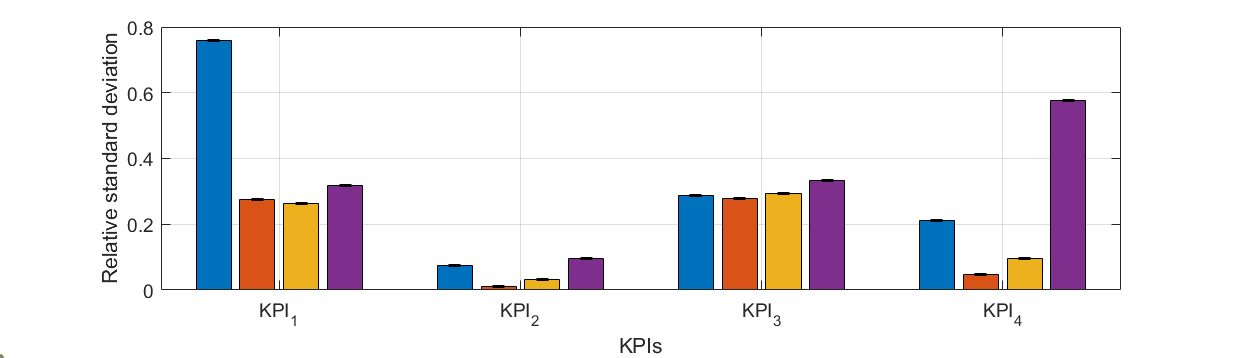}}
\caption{Ratio between the standard deviation and the mean (computed over 1000 samples) of the KPI values for the four considered configurations: \textcolor[RGB]{0, 114, 189}{$\blacksquare$} C1 $(\min J_1)$; \textcolor[RGB]{217, 83, 25}{$\blacksquare$} C2 $(\min J_2)$; \textcolor[RGB]{237, 177, 32}{$\blacksquare$} C3 $(\min J_3)$; \textcolor[RGB]{126, 47, 142}{$\blacksquare$} C4 $(\min J_4)$.}
\label{bars}
\end{figure*}

As discussed in Section~\ref{sec:gsa}, the GSA is aimed at evaluating the fraction of variance associated to the variation of an input parameter. In other words, it provides a ranking of the most influential input parameters but does not give information about the magnitude of the variance. For instance, an input parameter may be among the most influential, but since the total variance is small, the system would be sufficiently robust anyway, because the performance of the system would not anyway change significantly. In addition, the magnitude of the output variance should be related also to the average output values, since high variance values on low mean values could indicate a higher variability than, for instance, a high variance on very high mean values. Therefore, to fully evaluate the robustness of the PI framework, a preliminary analysis is performed in order to assess the ratio between the standard deviation and the mean (relative standard deviation in the following) of the above-mentioned KPIs. This measure is chosen as representative of the outputs variability. In this connection, Fig.~\ref{bars} shows the results, calculated with 1000 samples, for each KPI and configuration. 

As regards the results, $KPI_1$ (direct trucks usage) is characterized by a very high variability when it is minimized (configuration C1), and by a rather high variability when all the other objectives are minimized. Such a result indicates that variations in input parameters can significantly impact the usage of direct trucks.
As for $KPI_2$, i,e., the total delivery time of the PI-containers, its variability is significantly lower than the other KPIs, especially in configuration C2 and C3, indicating that the delivery time has a higher variability when the direct trucks usage or the delivery gap is minimized. 
The total transportation costs $KPI_3$ is characterized by high variability for all the configurations.
Lastly, $KPI_4$ is characterized by very high variability in configuration C4, when the total delivery gap is minimized, and high variability in configuration C1, when the direct truck usage is minimized.

By checking the effects of the different minimization goals (i.e., of the different configurations) on the KPIs, it is possible to note that minimizing direct truck usage (C1) leads to high variability in the model output on delivery time, total cost, and the delivery gap (blue bars in Fig.~\ref{bars}). Compared to the direct truck usage, minimizing the delivery time (orange bars) or the transportation costs (yellow bars) leads to lower variability of the model output, especially the delivery time.

In conclusions, from the analysis of the KPI variability, it is possible to conclude that the PI system results to be rather robust for the delivery time ($KPI_2$) with any optimization goal, i.e., in each configuration, and for the delivery gap ($KPI_4$) when the delivery time or the transportation costs are minimized.

\begin{figure*}[t!]
\centerline{\includegraphics[scale=0.50]{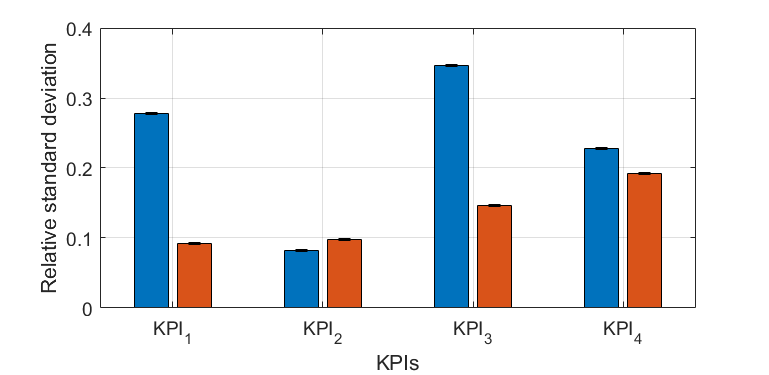}}
\caption{Ratio between the standard deviation and the mean (computed over 1000 samples, for configuration C2) by varying only the number of modules $n_s$ (\textcolor[RGB]{0, 114, 189}{$\blacksquare$}) or the processing time $\tau_p$ at one PI-hub (\textcolor[RGB]{217, 83, 25}{$\blacksquare$}).}
\label{bars2}
\end{figure*}

The above general analysis shows the variability of the model outputs when all the selected model inputs vary. Therefore, to isolate the effect of single inputs, the output variability was further investigated analyzing the results obtained when the number of modules per PI-container $n_s$ or the processing time $\tau_{p}$ of one PI-hub $p$ was free to vary at a time. This additional analysis was performed only for the configuration C2, since the total delivery time resulted to be, in the previous analysis, the one associated to the lowest output variability. The relevant results are reported in Fig.~\ref{bars2}, where the ratios between the standard deviation and the average values of the four KPIs are depicted. In such a figure, it is possible to note that all KPIs except $KPI_2$ are characterized by higher variability when $n_s$ vary and $\tau_p$ is fixed, with respect to the opposite case, showing the relevance of the number of PI-container modules in the PI framework optimized as in the model in Eq.~\eqref{eq:CF}--\eqref{25}, with respect to the processing times. Regarding $KPI_2$, it is worth underlying that the variability of the outputs is lower since it is the minimization goal.

\subsubsection{GSA results}
\label{res:gsa}

The last analysis performed on the considered PI network consists of the application of the GSA tool briefly recalled in Section~\ref{Sec:sensan}.
In this connection, based the results discussed in Section~\ref{variability_results}, the aim of this analysis consist of identifying the PI input parameters that are most responsible of such variability and to estimate the relevant sensitivity indices values. To this ain, the function $f$ mentioned in \label{Sec:sensan} is provided by the optimization model defined in in Eq.~\eqref{eq:CF}--\eqref{25} which dictates the considered PI framework. As regards input and outputs, the vector $X$ gathers the values $\tau_p$ of the operational times $\tau_p$ at the PI-hubs and the
number of modules $n_s$ of each PI-container and $Y$ consists of one of the KPIs $J_i$, $i=1,2,3,4$, at a time.

\begin{figure*}[htbp]
\centerline{\includegraphics[scale=0.55]{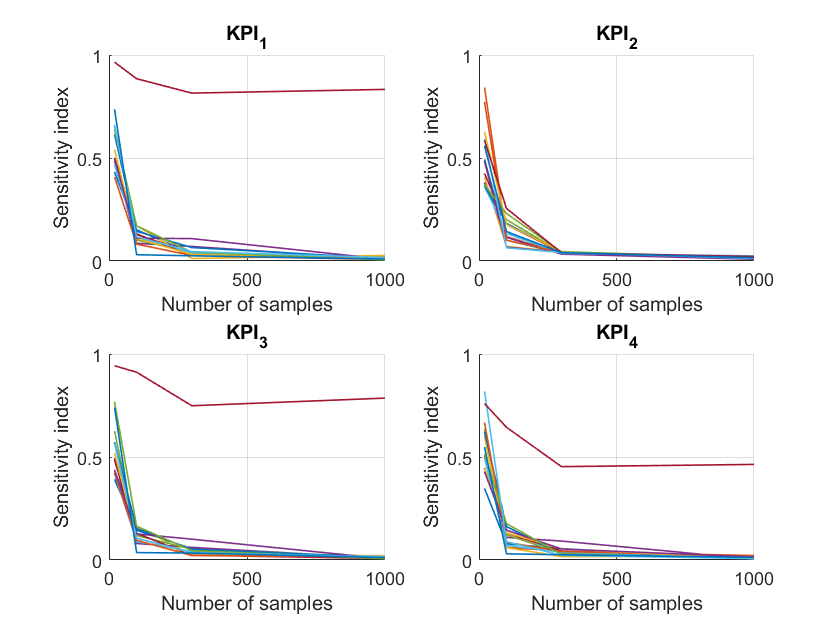}}
\caption{Sensitivity of the 15 inputs for the 4 KPIs in  configuration C2 as a function of the number of samples: \textcolor[RGB]{0, 114, 189}{$\blacksquare$} $\tau_1$; \textcolor[RGB]{217, 83, 25}{$\blacksquare$} $\tau_2$; \textcolor[RGB]{237, 177, 32}{$\blacksquare$} $\tau_3$; \textcolor[RGB]{126, 47, 142}{$\blacksquare$} $\tau_4$; \textcolor[RGB]{119, 172, 48}{$\blacksquare$} $\tau_5$; \textcolor[RGB]{77, 190, 238}{$\blacksquare$} $\tau_6$; \textcolor[RGB]{162, 20, 47}{$\blacksquare$} $\tau_7$; \textcolor[RGB]{0, 114, 189}{$\blacksquare$} $\tau_8$; \textcolor[RGB]{217, 83, 25}{$\blacksquare$} $\tau_9$; \textcolor[RGB]{237, 177, 32}{$\blacksquare$} $\tau_{10}$; \textcolor[RGB]{126, 47, 142}{$\blacksquare$} $\tau_{11}$; \textcolor[RGB]{119, 172, 48}{$\blacksquare$} $\tau_{12}$; \textcolor[RGB]{77, 190, 238}{$\blacksquare$} $\tau_{13}$; \textcolor[RGB]{162, 20, 47}{$\blacksquare$} $n_s$. 
}
\label{Sensindex}
\end{figure*}

The first step of this analysis consisted of identifying the number of samples needed to compute the sensitivity indices as it depends on the number of selected inputs, their probability distribution, and the model.
In this connection, consider the graph in Fig.~\ref{Sensindex}, where the value of the sensitivity indices are reported as a function of the number of samples. All the indices values have converged for 1000 samples, which was assumed sufficient for estimating the sensitivity indices values of all the input parameters.

\begin{figure*}[htbp]
\centerline{\includegraphics[scale=0.76]{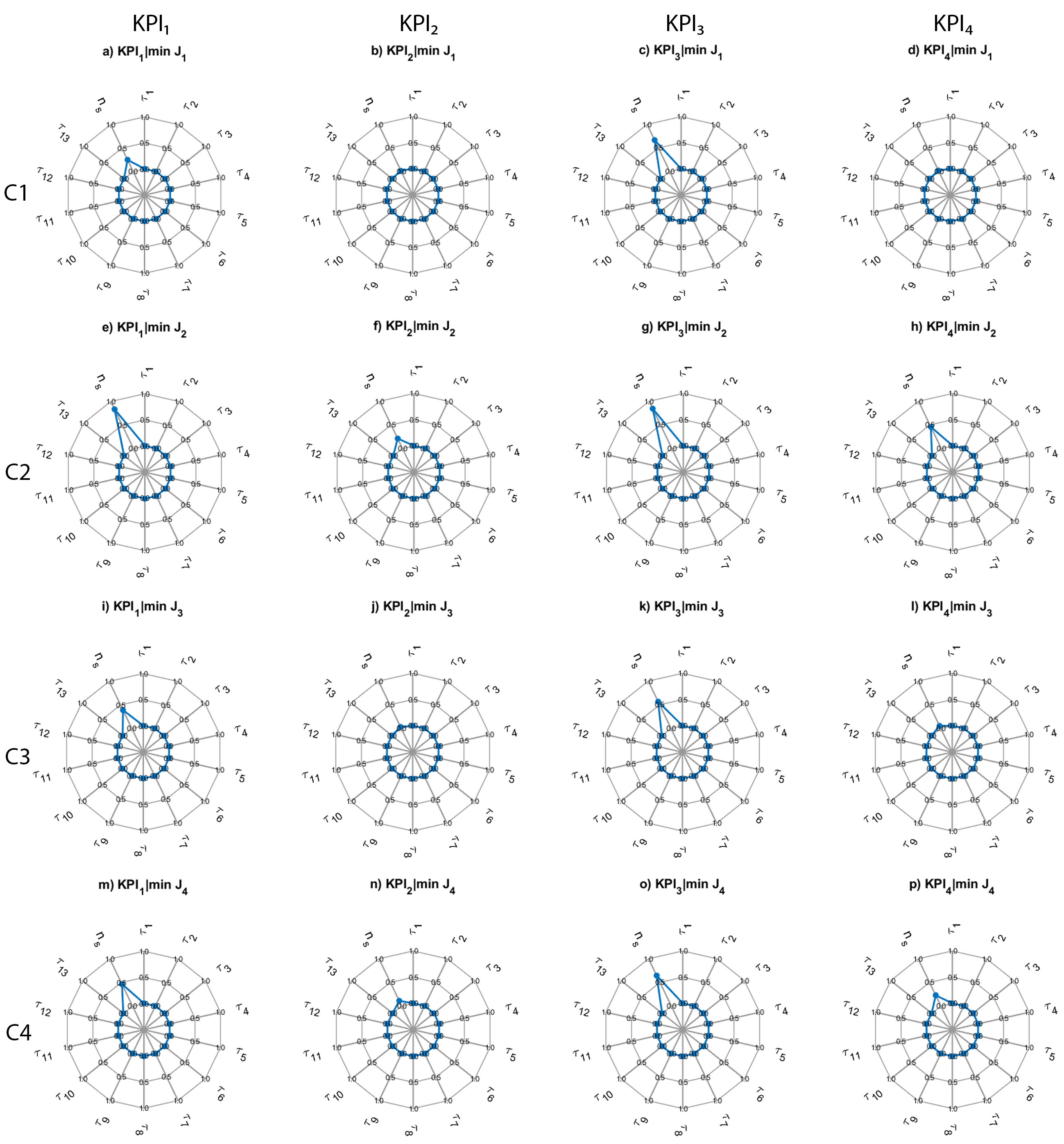}}
\caption{Sensitivity indices of the model inputs for the four KPIs (columns) and model configurations (rows).}
\label{spider}
\end{figure*}

Then, concerning the resulting sensitivity indices values, consider the radar graph reported in Fig.~\ref{spider} for the four configurations, addressed in the rows, and the four KPIs, addressed in the columns. 
In particular, attention is given to the cases where the PI system performance resulted to have more variability in the analysis performed in the previous section, that is, for $KPI_1$ (first column), $KPI_3$ (third column), and $KPI_4$ (fourth column), the latter only in configurations C1 and C4. In these cases, the sensitivity of the average number $n_s$ of modules per PI-container is significantly higher than the other ones. This indicates that such a parameter is relevant in determining the performance of the PI system, or, in other words, that the system is less robust with respect to such a parameter. In particular, $n_s$ is responsible for quite all the variability in the direct truck usage and transportation costs when the delivery time is minimized (Figs.~\ref{spider} e--g). Thus, this analysis suggests to consider it as a variable to be optimized by the model.

Conversely, the sensitivity values are significantly low for the operation times of the PI-hubs, indicating that the considered optimized PI scheme is robust with respect to an individual variation of such parameters, independently from the objective that is minimized. However, there are some cases (for example, $KPI_4$ for C1) where the sum of \textit{all} the indices is significantly lower than 1. This means that in such cases the performance variability is mainly caused by second or higher-order effects, i.e., by an interaction of the PI parameters.  
These interactions may imply that big variations of values of the KPIs are uniquely associated with specific combinations of the PI parameter values, in a way that cannot be caught by the computed first-order sensitivity indices. 
Anyway, from a practical point of view, this implies that a single PI-hub can be characterized by high variability in the processing time, without a significant consequence on the PI network performances. In other words, if the processing time of a single PI-hub is affected by higher uncertainty (e.g., due to the conditions in region/country), assuming its average in the whole network optimal programming would not significantly affect the whole logistic operations. 
On the other hand, if, for example, all the PI-hubs of the network were affected by high uncertainties, fixing all their processing times to the average values would not be recommended, as it is not possible to conclude that the PI system is sufficiently robust with respect to an interaction of them. Such a result suggests that even with a centralized optimization approach, a dynamic control approach to the module delivery should be considered, for instance by re-discussing the split and routing decisions every fixed time period to incorporate the most updated information about the problem inputs.

\section{Conclusions} 

In this paper, in order to better understand the PI framework and its related optimization model, a robustness analysis was conducted on the optimized PI logistic framework in order to assess the impact of the processing time of each PI-hub, and the number of modules in each PI-container. In this connection, after analyzing the variability of the system performance (described by four KPIs), caused by a change in the input parameter values, a sensitivity index, describing the relative influence of each input parameter on the model outputs, was computed.
According to the analysis of the system performance variability, the PI framework resulted rather robust (with respect to all the input parameters) when the system performance was evaluated on the basis of the total delivery time. This is a notable result, as the delivery time is one of the most common performance measures in logistics.
According to the GSA, the sensitivity indices of the processing times of each PI-hub were very low, indicating that the considered PI framework is quite robust to an individual variation of these parameters in all the KPIs considered.
On the contrary, the sensitivity index of the average number of modules in each PI-container was significantly greater than 0; this behavior suggests that this parameter should be considered relevant to the final solution obtained, and that the considered PI logistic scheme is less robust with respect to how the PI-containers are divided into modules. Hence, this information should be accurately considered in a real application and, possibly, included among the optimization variables.

The main limitation of this study is that, in the GSA, only the individual effect of each PI parameter (first-order effects) was evaluated. However, there are some cases (e.g., $KPI_4$ in C1, where the interactions (second or higher-order effects) among the PI parameters are prevalent, and thus it cannot be concluded that the model is robust with respect to these interactions. In these case, all the PI parameters should be treated as relevant, or further analyses should be conducted to evaluate the total effects.

Following the results found in this work, in future research, the way the PI-containers are divided into modules will be a decision variable of the optimization model, as this seems promising to further reduce the direct trucks usage and the transportation costs. Likewise, it can focus on implementing dynamic control strategies to periodically update split and routing decisions based on real-time system conditions. In addition, a heuristic solution approach will be developed to solve larger instances of the problem in a reasonable time. This could also allow to evaluate the total effects.

\section{Acknowledgements}

This work was supported by the European Union under the PNRR of NextGenerationEU, National  Sustainable Mobility Center (CN00000023, MUR Decree n. 1033—17/06/2022, Spoke 10). Views and opinions expressed are however those of the authors only and do not necessarily reflect those of the European Union or European Commission. Neither the European Union nor the European Commission can be held responsible for them.

\bibliographystyle{rusnat}
\bibliography{Mybib}
\end{document}